\documentstyle[12pt,psfig,amsfonts]{article}

\newcommand{\new}{\newcommand}

\new{\tnsr}{\otimes}                          %
\new{\comp}{\circ}                            %
\new{\NN}{{\Bbb N}}
\new{\CC}{{\Bbb C}}
\new{\RR}{{\Bbb R}}
\new{\FF}{{\Bbb F}}
\new{\ZZ}{{\Bbb Z}}
\new{\TT}{{\Bbb T}}

\new{\lbl}[1]{\label{#1}}

\newtheorem{proposition}{Proposition}
\newtheorem{theorem}{Theorem}
\newtheorem{lemma}{Lemma}
\newtheorem{claim}{Claim}
\newtheorem{corollary}{Corollary}
\newtheorem{conjecture}{Conjecture}

\newenvironment{pf}{
\begin{list}{\bf Proof:}{\vspace{12pt}
\setlength{\listparindent}{3em}
\setlength{\leftmargin}{.5in}\setlength{\rightmargin}{.2in}
}\item}{
\hfill \rule{1ex}{1ex}
\end{list}
\vspace{16pt}}

\newcounter{letter}
\newcounter{numeral}
\newcounter{Numeral}

\new{\prop}[1]{\begin{proposition} \lbl{#1}}
\new{\thm}[1]{\begin{theorem} \lbl{#1}}
\new{\lem}[1]{\begin{lemma} \lbl{#1}}
\new{\clm}[1]{\begin{claim} \lbl{#1}}
\new{\cor}[1]{\begin{corollary} \lbl{#1}}
\new{\cnj}[1]{\begin{conjecture} \lbl{#1}}
\new{\fig}{\begin{figure}[hbt]}
\new{\eq}[1]{\begin{equation} \lbl{#1}}

\new{\eprop}{\end{proposition}}
\new{\ethm}{\end{theorem}}
\new{\elem}{\end{lemma}}
\new{\eclm}{\end{claim}}
\new{\ecor}{\end{corollary}}
\new{\ecnj}{\end{conjecture}}
\new{\efig}[2]{\caption{#1} \label{#2} \end{figure}}
\new{\eeq}{\end{equation}}

\new{\pic}[5]{\raisebox{#3pt}{
\hspace{#4pt}\psfig{file=#1.ps,height=#2pt,silent=}\hspace{#5pt}}}

\begin{document}
\title{Path Integration in Two-Dimensional\\
Topological Quantum Field Theory}
\author{Stephen Sawin\\Dept. of Math\\MIT\\Cambridge, MA
  02139-4307\\sawin@math.mit.edu}

\maketitle
\thanks{This research supported in part by NSF postdoctoral Fellowship
       \#23068}
\begin{abstract}
A positive, diffeomorphism-invariant generalized measure on the space
of metrics of a two-dimensional smooth manifold is constructed. We use
the term generalized measure analogously with the generalized measures
of Ashtekar and Lewandowski and of Baez.   A
family of actions is presented which, when integrated against
this measure, give the two-dimensional axiomatic topological quantum
field theories, or TQFT's, in terms of which  Durhuus and Jonsson decompose
every
 two-dimensional unitary TQFT as a direct sum.
\end{abstract}

\section{Introduction}

This paper arose out of a question which the author first heard I. M. Singer
propose in a graduate course:  Does every two-dimensional axiomatic
topological quantum field theory  (TQFT) arise from a Lagrangian?  That is,
given Dijkgraaf's \cite{Dijkgraaf89}
classification of  two-dimensional TQFT's
\cite{Atiyah89,Atiyah90b}, can all such theories be formulated as the
path integral of some action?

Using the refinement of  Dijkgraaf's work due to Durhuus and Jonsson
\cite{DJ94}, this question is easy to answer on the physical level of
rigor.  They prove that each two-dimensional unitary TQFT is a direct
sum in an appropriate sense of theories arising from Euler number.
Specifically, each such theory is a direct sum of theories
${\cal Z}_\alpha$, for $\alpha \in \RR$, where ${\cal Z}_\alpha$
assigns a one-dimensional Hilbert space to the circle, and when
this is identified with $\CC$ correctly, assigns the number
$\exp(\alpha \chi(M))$ to every $2$-manifold
$M$ with boundary , where $\chi(M)$
is the Euler number of $M$.  Durhuus and Jonsson use the parameter
$\lambda= \exp(-\alpha/2)$ to parametrize their theories.

Such a theory admits a straightforward Lagrangian formulation.
The fields are metrics on $M$, and the Lagrangian is $\alpha/2\pi$
times the curvature two-form $\Omega_g$ of the metric $g$.  Of course,  if
$M$ is closed, then
$\int_M \Omega_g/2\pi=  \chi(M)$ by Gauss-Bonnet, and in
particular is a constant function on the space of metrics.  Thus
$${\cal Z}_\alpha(M)= e^{\alpha \chi(M)} =\int
e^{\alpha\int_M \Omega_g/2\pi}{\cal D}g.$$

If all we wanted to compute was the partition function of closed
manifolds, this would be perfectly rigorous and satisfactory.
However, it is clear one has not captured the notion of a path
integral satisfactorily if one cannot at least reproduce the cutting
and pasting arguments that justify Atiyah's axioms.  For this
one needs a space of fields and a measure on this space for
manifolds with boundary.

Here some difficulty occurs. As soon as one asks to be able to
integrate nontrivial functions, finding a diffeomorphism-invariant
measure on an infinite-dimensional space like the space of all metrics
becomes quite difficult, and perhaps impossible.  We  work instead
with `generalized measures', as developed by
Ashtekar and Lewandowski \cite{AL94,AL??} and Baez
\cite{Baez94,Baez94b}.  They work on the space of all connections of a
principal bundle, and take as their observables products of Wilson
loops.  They define a generalized measure on this space to be
something which ``knows the expectation values of these observables:''
i.e., a bounded linear functional on this $C^*$-algebra.  It turns out
that a generalized measure corresponds to an honest finite Borel
measure on an extension of the space of connections, called the space
of generalized connections.  Thus the whole theory of measure spaces
can be applied to such generalized measures.

This approach works well in our context.  A $2$-manifold has naturally
associated to it the principal $S^1$-bundle coming from the tangent
bundle, and a metric gives the Levi-Civita connection on this bundle.
A Wilson loop would just be $e^i$ raised to a multiple of the total geodesic
curvature of the loop.  It is more
convenient to deviate slightly from their formulation and take the
actual geodesic curvature, rather than the geodesic curvature mod
$2\pi$.  Thus our measurable functions will be limits of functions of
the total geodesic curvature of finitely many curves.  This
fortunately includes the action.

We construct a diffeomorphism-invariant generalized measure on the
space of metrics, with the
property that, for any countable set of curves, the total geodesic
curvature is almost always zero.  We will see that the TQFT's coming
from Euler number can be
made into rigorous path integrals against this measure.

This measure is an extremely simple one, and the theories that arise
from it certainly deserve the name toy models.  But, the geometric work
which must be done is probably   part of the work necessary
 to approach interesting generally covariant theories mathematically.
 In particular, this is a truly diffeomorphism-invariant theory.  That
 is, one
 can integrate functions of the curvatures of any countable collection
 of smooth curves, even if the curves are not  analytic.  It is hoped
 that this will be a step towards constructing fully
 diffeomorphism-invariant generalized measures on the space of
 connections with a nonabelian group.

This paper is not the first to give rigorous path-integral
formulations of TQFT's.  \cite{FHK94} gives a state-sum on a
triangulation for theories ${\cal Z}_\alpha$, when $\exp(\alpha) \in
\NN$, and \cite{FQ93} describes theories  based on finite-groups, which are
a subset of these, as a finite
sum.  In both cases the sums may be taken as discrete path integrals,
and can be fit into the framework of Section III.

I would like to thank Isadore Singer,  John Baez, Scott Axelrod, Eric
Weinstein, Dan Stroock and Richard Dudley for comments, suggestions  and
conversations.

\section{Constructing the Generalized Measure}

Let $M$ be a smooth, compact $2$-manifold, possibly with boundary.
By a curve in $M$ we will always mean a  smooth immersion of the circle
into $M$,
considered up to positive reparametrization.

If $c:S^1 \to M$ is a smooth immersion, the geodesic
curvature of $c$ with respect to a metric $g$ at a point $t \in S^1$,
$\kappa_{c,g}(t)$, is the value of the Levi-Civita connection one-form
of $g$ on the tangent to the lift of the curve to $TM$, at $t$.  It depends on
the parametrization, but the total curvature, $k_c(g)=\int_{S^1}
\kappa_{c,g}(t)
dt$, does not, and thus is a function of the curve $c$.

If $c_1, \ldots,c_n$ are curves in $M$, and $f$
is any bounded continuous function on  $n$ real variables, we get a
bounded continuous function $F_{f,c_1,\ldots,c_n}$ on ${\cal G}$, the
set of all metrics on
$M$, sending $g \in {\cal G}$ to
$f(k_{c_1}(g), \ldots,k_{c_n}(g))$.  Continuous bounded complex-valued
functions on
${\cal G}$ form a commutative $C^*$-algebra with complex
conjugation as involution and the sup norm as norm, and the closure
$C$ in this norm of the algebra of all functions
$F_{f,c_1,\ldots,c_n}$ is
a $C^*$-subalgebra. We will call a
bounded linear functional on $C$ a {\em generalized
  measure on ${\cal G}$\/}.

The group of diffeomorphisms of $M$ acts as a group of homeomorphisms
on ${\cal G}$, and thus as a group of $C^*$-isomorphisms on the
bounded continuous functions on  ${\cal G}$.  These isomorphisms take $C$ to
itself, so they also act on $C$, where a diffeomorphism $D$ sends
$F_{f,c_1,\ldots,c_n}$ to $F_{f,D(c_1),\ldots,D(c_n)}$.  Call a
generalized measure {\em diffeomorphism-invariant\/} if its value on
an element of $C$ is unchanged by the action of a diffeomorphism.
Also, call a generalized measure {\em positive\/} if its value on any
nonnegative-valued function in $C$ is nonnegative.

The terminology warrants an explanation.  By a fundamental  theorem of
$C^*$-algebras
\cite[Thm. 1.4.1]{Dixmier77}, $C$ is
the algebra of continuous functions on some compact Hausdorff space
${\frak G}$ into which
 ${\cal G}$  maps continuously and densely.  In
particular,  a bounded linear functional on $C$ corresponds by the
Riesz representation theorem to a finite Borel measure on
${\frak G}.$  What's more the action of  diffeomorphisms of $M$ on
${\cal G}$ extends to a continuous action   on ${\frak G}$,
and thus acts on the finite Borel measures of ${\frak G}$, the
action being the one already described for linear functionals on $C$.
Thus diffeomorphism-invariant generalized measures on ${\cal G}$ are
exactly diffeomorphism-in\-var\-i\-ant finite Borel measures on ${\frak G}$.
Likewise, an element of $C$ is nonnegative if and only if it is nonnegative as
a
function on ${\frak G}$, so positive generalized measures are exactly
positive, finite, Borel measures on ${\frak G}$.

This is the analogue of the notion of
generalized measures on the space of
connections developed  by Ashtekar \& Lewandowski and Baez.  It is
helpful to think of the generalized measure as an honest measure on
the enlarged space, and to rely on our knowledge and intuition about
measures.  This is what we do in reconstructing the TQFT from the
generalized measure we construct.  It would be possible to do
everything without reference to ${\frak G}$, simply relying on
the definition of a generalized measure as a bounded linear
functional.   This certainly has an appealing concreteness, and might
dispel a slightly mystical feel associated to these highly abstract
spaces and structures.  The language of path integrals is so
compelling, however, that the exposition seems greatly aided by the
ability to use explicit measures and integrals.

The measure we construct is defined by the very simple property that
for any finite set of  curves, almost every generalized metric assigns
these curves total curvature zero.

\thm{th:measure-existence} There exists a positive, diffeomorphism-invariant
generalized
  measure sending each $F_{f,c_1, \ldots,c_n}$ to
  $f(0, \ldots, 0)$.
\ethm

To prove the functional is bounded, in fact of norm one, the key
geometric step is to see that its value on a function is contained in
the range of the function: that for any finite set of curves,
there is a metric which makes their total curvatures all
simultaneously zero.  This straightforward sounding fact is quite
subtle, because of the complicated manner in which the curves might intersect.

\lem{lm:flat} Let $\{c_i\}_{i=1}^n$ be a  set of
curves in a smooth $2$-manifold $M$.
There
is a metric on $M$  which makes the total curvature
on each $c_i$ equal to zero.
\elem

\begin{figure}[h]
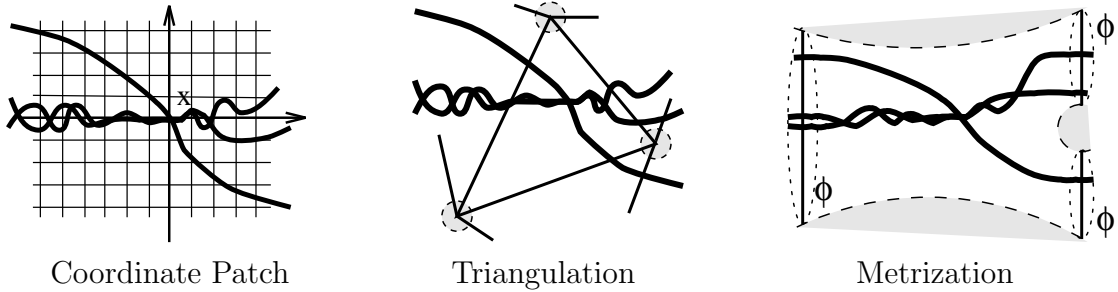

\begin{tabular*}{5.9in}{ccc}
\pic{patch}{90}{0}{0}{0} \hspace{8pt} &
\hspace{8pt} \pic{triangle}{90}{0}{0}{0} \hspace{8pt} &
\hspace{8pt} \pic{metrize}{90}{0}{0}{0}\\
Coordinate Patch & Triangulation & Metrization
\end{tabular*}
\caption{Putting a flat metric on a piece of $K$}
\label{fg:triangleexample}
\end{figure}

\begin{pf}  First, recall
that no curve can touch the same point $x$ in $M$ infinitely many times.  If
it did, the preimage of $x$ in $S^1$ would have an accumulation point,
and the curve would not be an immersion at that point.

Consider each $x\in K$, the union of the ranges of the $c_i$'s.
  Identify the tangent plane to $M$ at $x$ with $\RR^2$ in such a fashion that
no
  tangent vector to any $c_i$ is on the $y$-axis.  Extend this to a
  coordinate patch, and consider a neighborhood $N_x$ small enough that no
  tangent to $c_i$ is  parallel to the $y$-axis.  Since $K$ is
  compact, it is covered by finitely many such  $N_i$'s, $i=1, \ldots,
  k$.  Choose a triangulation of $M$, and subdivide as necessary until
  every triangle intersecting $K$ lies in some $N_i$.  Perturb
  this triangulation  slightly
  so that the vertices do not lie on $K$ and no edge is ever
  horizontal with respect to any coordinate patch.  Notice each $c_i$
  intersects each edge
  transversely, and hence finitely many times.

Now choose
  nonintersecting neighborhoods of
  each edge $e$ minus a neighborhood of the vertices, and identify them
   with a region of $\RR^2$ via a map $\phi_e$, so that the edge is
   identified with a straight line and each intersection with a $c_i$
   is perpendicular.  Finally, identify each open face of the
   triangulation  minus a disk around each vertex with a region
   of $\RR^2$, so that the overlap maps with $\phi_e$ are isometries,
   and all three edges are vertical, pointing up or down according to
   whether they pointed up or down in the original coordinate patch.
   See Figure \ref{fg:triangleexample} for an example of this process.

The metric induced by these maps is a flat metric on
$M$ with  finitely many disks removed.  Extend it to all of $M$.  The total
curvature of $c_i$ with respect to this metric is the sum of the total
curvature of $c_i$ in each triangle.  However, within one triangle $c_i$
goes without
self-intersection from being horizontal to being horizontal with the same
orientation.  Since the metric is flat, the total curvature of
this
piece is zero.  Thus the total curvature of each $c_i$ in this metric
is zero.
\end{pf}

\begin{pf} {\em (Of Theorem \ref{th:measure-existence})\/}  It
  suffices to show this gives a well-defined norm-one linear
  functional on all functions of the form $F_{f,c_1,\ldots,c_n}$,
  since these are dense in $C$.  If $F_{f,c_1,\ldots,c_n}$ and
  $F_{g,d_1,\ldots,d_n}$ represent the same function, find  by Lemma
  \ref{lm:flat} a metric such that all $c_i$ and $d_i$ have zero total
  curvature.  In each case the value of the functional is just the
  value of the function on this metric, and therefore is the same.  Since the
value of the
  functional is always gotten by evaluation of the function at some
  point, it is clearly smaller than the sup norm of that function, and
  therefore is of norm one.
\end{pf}

Call the measure we have constructed ${\cal D}g$.

\section{Constructing the TQFT}

We must say what it means for a TQFT to `arise from a path integral.'

Consider a contravariant functor from the tensor category of
smooth, compact $n$-manifolds with boundary and smooth
immersions
to the tensor category of probability
spaces and measurable functions.
Specifically, this is an assignment of a probability space ${\frak F}_M$
to every manifold $M$, and a measurable map
${\frak F}_f:{\frak F}_N  \to{\frak F}_M$ to each immersion $f:M \to N$, such
that
${\frak F}_g {\frak F}_f = {\frak F}_{fg}$, ${\frak F}_1=1$, ${\frak F}_{M
  \cup N}={\frak F}_M \times {\frak F}_N$, ${\frak F}_{f
  \cup g}={\frak F}_f \times {\frak F}_g$, the empty $n$-manifold gets
sent to a one-point probability space, and the obvious map from $M
\cup N$ to $N \cup M$ gets sent to the obvious map from ${\frak F}_M
\times {\frak F}_N$ to ${\frak F}_N
\times {\frak F}_M$.  We also ask that if $i$ is an embedding then
${\frak F}_i$ is measure-preserving, in the sense that the preimage of
measurable sets have the same measure.  Thus in particular
diffeomorphisms correspond to measure-space isomorphisms.

If $i:N \to M$ is an embedding, so that ${\frak F}_i:{\frak F}_M \to
{\frak F}_N$ is measure-preserving,  consider an integrable function $f$
from ${\frak F}_M$ to $\CC$.  It defines a measure on ${\frak F}_N$, by
assigning to each measurable set the integral of $f$ on the pullback
of that set.  This measure is absolutely continuous with respect to
the probability measure on ${\frak F}_N$, and thus arises from  some
function $E_N^M(f)$ defined a.e., by the Radon-Nikodym theorem.   The map
$E$ is called a conditional expectation, and satisfies that, for
$f$, $f'$ functions on ${\frak F}_M$ and ${\frak F}_N$ respectively,
$$\int_{{\frak F}_M} E^M_N(f) f' {\cal D}g= \int_{{\frak F}_M} ff'
{\cal D}g,$$
where all functions are pulled back to ${\frak F}_M$.

  Our final condition on this
functor is that if $M_1 \cap M_2 \subset N$ are all submanifolds of
some $M$, $f_1$ and $f_2$ are integrable functions on
${\frak F}_{M_1}$ and ${\frak F}_{M_2}$ respectively, and the product of
their pullbacks to ${\frak F}_M$ is also integrable, then
$E^M_{N}(f_1f_2)=E^M_N(f_1)E^M_N(f_2)$, where by $f_1$ and $f_2$ we
really mean their pullbacks to ${\frak F}_M$.  This odd condition
expresses the intuitive notion that having specified a field on $N$,
the possible ways of extending it to $M_1$ and  $M_2$ are independent.

Suppose a portion of the boundary of $M$ has been identified with a
$d-1$-manifold $\Sigma$ in some way.  This identification can be
extended to an embedding $i$ of $\Sigma \times I$ into $M$.  The
embedding is certainly not unique, but two different such can be
related by a boundary-fixing automorphism of $M$, so that the
measure-preserving map ${\frak F}_i:{\frak F}_M \to {\frak F}_{\Sigma
  \times I}$ is
unique up to a measure-isomorphism of ${\frak F}_M$ (again, having picked
an identification of the boundary with $\Sigma$).  If we define
${\frak F}_\Sigma$ to be ${\frak F}_{\Sigma \times I}$,
${\frak F}$ is  a contravariant functor with the same properties
for $n-1$-manifolds, and there is a measure preserving natural
transformation ${\frak F}_M \to {\frak F}_{\partial M}$.

Now for each compact, closed,  {\em oriented\/} $n$-manifold $M$, let
$S_M$ be an integrable function on ${\frak F}_M$.  Suppose  $S_{M\cup N}= S_M
\times S_N$, $S_M = \bar{S}_{M^*}$, the bar indicating complex
conjugate and $M^*$ indicating $M$ with the reverse orientation, and
if $f:M \to N$ is an onto immersion, then
$S_N=S_M{\frak F}_f$.
Call such an $S$ an {\em
  action\/}.  Define ${\cal Z}(M)= E^M_{\partial M}(S_M)$, an
integrable function on
${\frak F}_{\partial M}$.  Notice that ${\cal Z}(M)=
\overline{{\cal Z}(M^*)}$.

Let $M_1$ have a portion of its boundary isomorphic to an
$n-1$-manifold $\Sigma$, and let $M_2$ have a portion of its boundary
isomorphic to $\Sigma^*$, and consider the manifold $M$ formed by
gluing them together by identifying points along this boundary.  We
may just as well identify a neighborhood of both boundaries with
$\Sigma\times I$, and identify them along these submanifolds.   Thus
if $\Sigma_1$ and
$\Sigma_2$ are the rest of the boundary of $M_1$ and $M_2$
respectively, we have
\begin{eqnarray*}
E^M_{\Sigma_1 \cup \Sigma_2}(S_M)&=& E^M_{\Sigma_1 \cup
  \Sigma_2}(S_{M_1}S_{M_2}) \\
&=& E^{\Sigma_1 \cup \Sigma_2 \cup \Sigma}_{\Sigma_1 \cup
  \Sigma_2} E^M_{\Sigma_1 \cup \Sigma_2 \cup
  \Sigma}(S_{M_1}S_{M_2})\\
&=&E^{\Sigma_1 \cup \Sigma_2 \cup \Sigma}_{\Sigma_1 \cup
  \Sigma_2} E^{M}_{\Sigma_1 \cup \Sigma \cup \Sigma_2}(S_{M_1})
E^{M}_{\Sigma_2 \cup \Sigma \cup \Sigma_1}(S_{M_2})\\
&=&E^{\Sigma_1 \cup \Sigma_2 \cup \Sigma}_{\Sigma_1 \cup
  \Sigma_2} E^{M_1}_{\partial{M_1}}(S_{M_1})E^{M_2}_{\partial{M_2}}(S_{M_2})
\end{eqnarray*}
with $d-1$-manifolds such as $\Sigma_1$ used to denote the associated
subspaces of the $d$-manifolds diffeomorphic to $\Sigma_1 \times I$,
for convenience.  The last step is because $\Sigma_2$ is disjoint from
$M_1$ and $\Sigma \cup \Sigma_1$, so the probability spaces of the unions
are products, and thus $E^{M}_{\Sigma_1 \cup \Sigma \cup \Sigma_2}=
E^{M_1 \cup \Sigma_2}_{\Sigma_1 \cup \Sigma \cup
  \Sigma_2}=E^{M_1}_{\Sigma_1 \cup \Sigma} \times {\rm id}$
for functions of ${\frak F}_{M_1} \times {\frak F}_{\Sigma_2}$.
Now ${\frak F}_{\Sigma_1 \cup
  \Sigma_2 \cup \Sigma}$
is a product probability space, so denoting elements of
${\frak F}_{\Sigma_1}$, ${\frak F}_{\Sigma_2}$, and ${\frak F}_{\Sigma}$
by $a$, $b$, and $c$ respectively, this says
$${\cal Z}(M)(a,b)= \int {\cal Z}(M_1)(a,c) {\cal Z}(M_2)(b,c)
{\cal D}c.$$

In particular, gluing $M$ to $M^*$ along their common boundary and
taking ${\cal Z}$ of the result yields
$\int {\cal Z}(M) \overline{{\cal Z}(M)}$, and we see ${\cal Z}(M)$ is in
$L_2({\frak F}_{\partial M})$ for all $M$.  We may thus interpret
${\cal Z}(M)$, ${\cal Z}(M_1)$ and ${\cal Z}(M_2)$ above as the kernels
of Hilbert-Schmidt operators ${\frak F}_{\Sigma_1} \to
{\frak F}_{\Sigma_2}$, ${\frak F}_{\Sigma_1} \to
{\frak F}_{\Sigma}$, and ${\frak F}_{\Sigma} \to
{\frak F}_{\Sigma_2}$ respectively, and then the above gluing law
becomes
$${\cal Z}(M)= {\cal Z}(M_2) {\cal Z}(M_1)$$
the product being composition of Hilbert-Schmidt operators.

Now consider ${\cal Z}(\Sigma \times I) \in L_2({\frak F}_\Sigma \times
{\frak F}_{\Sigma})$ as the kernel of a
Hilbert-Schmidt operator $P_\Sigma$ on
$L_2({\frak F}_{\Sigma})$.  Let ${\cal Z}(\Sigma)$ be the
range of $P_\Sigma$.

\thm{th:TQFT}
Let ${\frak F}$ be a contravariant functor as above, and let $S$ be an
action for it.
Then ${\cal Z}$ is a unitary TQFT.
\ethm

\begin{pf}
We first argue that ${\cal Z}(M)$ lies in ${\cal Z}(\partial M)$.
Notice that $M$ is equal to $M$ glued to $\partial M \times I$.  In
particular,  ${\cal Z}(M) = P_{\partial M}{\cal Z}(M)$ and thus is in
${\cal Z}(\Sigma)$.

We have already demonstrated the gluing law.  In particular
$P_\Sigma P_\Sigma$ is the operator associated to the manifold
formed by gluing $\Sigma \times I$ to itself along one $\Sigma$.
Since this is $\Sigma \times I$ again, We have $P_\Sigma P_\Sigma
= P_\Sigma$, and thus it acts as the identity on its range
${\cal Z}(\Sigma)$.

Since  ${\cal Z}(\Sigma_1 \cup \Sigma_2)$ is the range of  $P_{\Sigma_1
  \cup \Sigma_2} = P_{\Sigma_1} \tnsr P_{\Sigma_2}$, it is the
tensor product of the ranges of the two factors, so it is
${\cal Z}(\Sigma_1) \tnsr {\cal Z}(\Sigma_2)$.  Of course, if
$\Sigma$ is the empty $n-1$ manifold then ${\frak F}_\Sigma$ is
the one-point probability space and $P_\Sigma$ is the identity, so
${\cal Z}(\Sigma)= \CC$.

Now $P_\Sigma$ is Hilbert-Schmidt and hence compact, so acting as
the identity on ${\cal Z}(\Sigma)$ means that ${\cal Z}(\Sigma)$ is
finite-dimensional. Thus we can write ${\cal Z}(\Sigma \times I)$
as $\sum_{i=1}^n f_i \tnsr \bar{f}_i$, where $\{ f_i\}$ is an
orthonormal basis for ${\cal Z}(\Sigma)$ of functions of
${\frak F}_\Sigma$.  Since ${\cal Z}(\Sigma \times I)$ is also an
element of ${\cal Z}(\Sigma) \tnsr {\cal Z}(\Sigma^*)$, we have
that ${\cal Z}(\Sigma^*)$ contains the complex conjugate of
${\cal Z}(\Sigma)$ in $L_2({\frak F}_\Sigma)$.  Since
$(\Sigma^*)^*= \Sigma$, ${\cal Z}(\Sigma^*)$ is equal to the complex
conjugate of ${\cal Z}(\Sigma)$, and
thus ${\cal Z}(\Sigma^*)$ is dual in the obvious way to
${\cal Z}(\Sigma)$. ${\cal Z}(\Sigma \times I)$ is the canonical element
of ${\cal Z}(\Sigma) \tnsr {\cal Z}(\Sigma)^*$.

We have already seen that any diffeomorphism of $\Sigma$ to
$\Sigma'$ corresponds to a measure space isomorphism of
${\frak F}_\Sigma$ to ${\frak F}_{\Sigma'}$.  This diffeomorphism
extends to a diffeomorphism of $\Sigma \times I$ to $\Sigma'
\times I$, and so the isomorphism takes ${\cal Z}(\Sigma \times
I)$ to ${\cal Z}(\Sigma' \times I)$.  We thus get an isomorphism
from ${\cal Z}(\Sigma)$ to ${\cal Z}(\Sigma')$.  More generally, if
$M$ is diffeomorphic to $M'$, the diffeomorphism restricted to
the boundary gives an isomorphism of ${\cal Z}(\partial M)$ to
${\cal Z}(\partial M')$ sending ${\cal Z}(M)$ to ${\cal Z}(M')$.
This and the functoriality of ${\frak F}$ make ${\cal Z}$
functorial.

Finally, since ${\cal Z}(M^*)= \overline{{\cal Z}(M)}$, we have
that they are the  kernels of adjoint Hilbert-Schmidt operators.
This is all we needed to show for ${\cal Z}$ to be a unitary TQFT
\cite{Atiyah89,Atiyah90b}.
\end{pf}

\thm{th:Euler}  The assignment to each $M$ of the measure space ${\frak G}$ of
generalized measures on $M$ with the measure constructed in the previous
section, together with the action $S_M= \exp(\alpha \int_M
\Omega_g/2\pi)$, for a fixed $\alpha \in \RR$,  satisfy the
assumptions of the previous theorem,
and thus give a $2$-dimensional unitary TQFT. What's more, this
is exactly the TQFT ${\cal Z}_\alpha$ described by Durhuus and
Jonsson \cite{DJ94}.
\ethm
\begin{pf}
We have not fully described the data, since we have not
associated a measurable map to each immersion.  Let $i:M
\to N$ be such an immersion.  This gives a map $i^*: {\cal G}_N
\to {\cal G}_M$, taking each metric on $N$ to its pullback on $M$.
If $F_{f,c_1,\ldots , c_n}$ is a typical element of $C$ for $M$,
then $F_{f,c_1,\ldots c_n}\comp i^*= F_{f,i(c_1), \ldots,
  i(c_n)}$, an element of $C$ for $N$.  This is a
$C^*$-homomorphism, and thus induces a
continuous onto  map  ${\frak G}_i:{\frak G}_N \to {\frak G}_M$. If $i$
is an embedding, the integral of each $F$ is unchanged by ${\frak G}_i$,
so it is measure-preserving.  The assignment of $i^*$ to $i$ is
a contravariant functor, so
the assignment ${\frak G}$ is
too.

The conditional expectation $E^M_N$ is trivial, because the space of
integrable functions up to a.e. equivalence is one dimensional.  In
fact, if $N$ is nonempty$E^M_N(F_{f,c_1, \ldots, c_n})=F_{g,d_1,
  \ldots, d_k}$, where
$d_i$ are any curves on $N$ and $g$ is any function with
$g(0,\ldots,0)=f(0,\ldots,0)$.  If $N$ is the empty manifold, then it
is the function on the one-point probability space with value
$f(0,\ldots, 0)$.  Either way it is always true that
$E^M_N(f_1)E^M_N(f_2)=E^M_N(f_1f_2)$ a.e., for any functions $f_i$ on
${\frak F}_M$.

$S_M$ is an element of $C$, because it is just $\alpha( \chi(M) +
x_g/2\pi)$, where $x_g$ is the total curvature around the
boundary.  Thus it is integrable. It clearly pulls back through onto immersion,
its value on a
union of manifolds is the product of its values on the individual
manifolds, and reversing orientation does not change it and thus
in particular sends it to its complex conjugate for $\alpha \in \RR$.

The TQFT is now easy to describe.  An element of ${\frak G}_M$
assigns a total curvature to each curve, and two will go to the
same element of ${\frak F}_{ M'}$ if they assign the same
value to all curves in $M'$.  $L_2$ functions on
${\frak G}_{ M'}$ will be spanned by those of the form
$F_{f,c_1, \ldots, c_n}$ with the $c_i$'s in $ M'$.  Such a
function has zero norm if $f(0,\ldots,0)=0$.  Thus
$L_2({\frak G}_\Sigma)= \CC$ for every one-manifold $\Sigma$.  For
any $M$, ${\cal Z}(M)$ is the function $\exp(\alpha (\chi(M) +
x/2\pi))$,  where $x$ is the total curvature around the boundary,
and thus corresponds as an element of $\CC$ to $\exp(\alpha
\chi(M))$.  Each $P_\Sigma$ is then $1$, so ${\cal Z}(\Sigma)=
\CC$, with ${\cal Z}(M)= \exp(\alpha \chi(M))$.  This is exactly
the  TQFT constructed by Durhuus and Jonsson.
\end{pf}

\section{Remarks}
\begin{itemize}

\item All that we did here would work as well for piecewise
  smooth curves, and for curves with endpoints.

\item Of course, these theories have obvious observables,
  namely  functions of the total curvature of a given set
  of closed curves.  Unfortunately, the expectation value of these
  observables are all zero.

\item
One might wonder whether the work of constructing a generalized
measure was really necessary.  If we had
defined the $C^*$-algebra $C$ to be spanned only by functions of the
total curvature of the boundary of $M$,  we would have had no
difficulty finding a measure on ${\cal G}$ which made everything in $C$
integrable and satisfied the restrictions necessary to make the value
of the partition function ${\cal Z}_\alpha$ (specifically, the integral
of $\exp((\alpha+\bar{\alpha})x_B/2\pi)$ must be $1$, where $x_B$ was the total
curvature of the boundary).  However, such a measure would not
have had the functorial properties needed to apply Theorem
\ref{th:TQFT}.  By reproducing in a rigorous fashion the
heuristic arguments leading from path integrals to Atiyah's
axioms, Theorem \ref{th:TQFT} represents a
reasonable definition of what it means for a TQFT to arise
rigorously from a path integral.

\item
We began with the question: Do all two-dimensional axiomatic
unitary topological quantum field theories arise as path integrals?
We have only actually shown that every such is a direct sum of
theories with this property.  But it is straightforward
to write the direct sum of two path integral theories as a path
integral: for connected $M$ the measure space of fields is just the
union of the two measure spaces, and the action is just the function
which is each respective action on each component.

\item
A slight modification of the above argument gives a path-integral
formulation of the nonunitary TQFT ${\cal Z}_\alpha$ for $\alpha \in \CC- \RR$.
These represent many of the
irreducible nonunitary two-dimensional TQFT's, but not all:  there are
nilpotent theories \cite{Sawin2D} for which no path integral
formulation is known.  They are simple enough that one might
conceivably hope to
find actions for them, but their geometry is less transparent than the
${\cal Z}_\alpha$.
\end{itemize}

\bibliography{/m3/sawin/P/references}

\begin{thebibliography}{Bae94b}

\bibitem[AL]{AL??}
A.~Ashtekar and J.~Lewandowski.
\newblock Projective techniques and functional integration for gauge theories.
\newblock to appear in J. Math. Phys.

\bibitem[AL94]{AL94}
A.~Ashtekar and J.~Lewandowski.
\newblock Representation theory of analytic holonomy {$C^*$}-algebras in knots
  and quantum gravity.
\newblock In J.~Baez, editor, {\em Knots and Quantum Gravity}, Oxford, 1994.
  Oxford U. Press.

\bibitem[Ati89]{Atiyah89}
M.~F. Atiyah.
\newblock Topological quantum field theories.
\newblock {\em Publ. Math. IHES}, 68:175--186, 1989.

\bibitem[Ati90]{Atiyah90b}
M.~F. Atiyah.
\newblock {\em The Geometry and Physics of Knots}.
\newblock Lezioni Lincee. Cambridge University Press, 1990.

\bibitem[Bae94a]{Baez94b}
J.~Baez.
\newblock Diffeomorphism-invariant generalized measures on the space of
  connections modulo gauge transformations.
\newblock In D.~Yetter, editor, {\em Proceedings of the Conference on Quantum
  Topology}, Singapore, 1994. World Scientific.

\bibitem[Bae94b]{Baez94}
J.~Baez.
\newblock Generalized measures in gauge theory.
\newblock {\em Lett. Math. Phys.}, 31:213--223, 1994.

\bibitem[Dij89]{Dijkgraaf89}
R.~H. Dijkgraaf.
\newblock {\em A Geometric Approach To Two-Dimensional Conformal Field Theory}.
\newblock PhD thesis, University of Utrecht, 1989.

\bibitem[Dix77]{Dixmier77}
J.~Dixmier.
\newblock {\em $C^*$-Algebras}.
\newblock North Holland Publishing Company, Amsterdam-New York-Oxford, 1977.

\bibitem[DJ94]{DJ94}
B.~Durhuus and T.~Jonsson.
\newblock Classification and construction of unitary topological quantum field
  theories in two dimensions.
\newblock {\em J. Math. Phys.}, 35(10):5306--5313, October 1994.

\bibitem[FHK94]{FHK94}
M.~Fukuma, S.~Hosono, and H.~Kawai.
\newblock Lattice topological field theory in two dimensions.
\newblock {\em Comm. Math. Phys.}, 161(1):157--175, 1994.

\bibitem[FQ93]{FQ93}
D.~Freed and F.~Quinn.
\newblock {Chern-Simons} theory with finite gauge group.
\newblock {\em Comm. Math. Phys.}, 156:435--472, 1993.

\bibitem[Saw]{Sawin2D}
S.~Sawin.
\newblock Direct sum decompositions and irreducible {TQFT}'s.
\newblock preprint.

\end{thebibliography}
\bibliographystyle{alpha}

\end{document}